\documentclass[aps,preprint,prb,groupedaddress,showpacs]{revtex4}
\usepackage{graphicx}
\def\sigmav{{\mbox{\boldmath{$\sigma$}}}}

\def\deltav{{\mbox{\boldmath{$\delta$}}}}
\begin{document}
\bibliographystyle{apsrev}
%

\title{HIDDEN SYMMETRIES  AND THEIR CONSEQUENCES 
IN $t_{2g}$ CUBIC PEROVSKITES}
\author{A. B. Harris$^{1}$, T. Yildirim$^{2}$,
A. Aharony$^{3}$, O. Entin-Wohlman$^{3}$, and I. Ya. Korenblit$^3$}
\affiliation{$^1$Department of Physics and Astronomy, University of
Pennsylvania, Philadelphia, PA 19104, \\
$^{2}$NIST Center for Neutron Research,
National Institute of Standards and Technology, Gaithersburg, MD
20899 \\
$^{3}$School of Physics and Astronomy, Raymond and Beverly Sackler
Faculty of Exact Sciences,  Tel Aviv University, Tel Aviv 69978,
Israel}

\date{\today}
\begin{abstract}

The five-band Hubbard model for a $d$ band with one electron
per site is a model which has very interesting properties
when the relevant ions are located at sites with high
(e. g. cubic) symmetry.  In that case, if the crystal
field splitting is large one may consider excitations confined
to the lowest threefold degenerate $t_{2g}$ orbital states.
When the electron hopping matrix element ($t$) is much
smaller than the on-site Coulomb interaction energy ($U$),
the Hubbard model can be mapped onto the well-known effective
Hamiltonian (at order  $t^{2}/U$) derived by Kugel and Khomskii (KK).  
Recently we have shown that the KK Hamiltonian does not support
long range spin order at any nonzero temperature
due to several novel hidden symmetries that it possesses.
Here we extend our theory to show that these symmetries
also apply to the underlying three-band Hubbard model.
Using these symmetries we develop a rigorous Mermin-Wagner
construction, which shows that the three-band Hubbard model
does not support spontaneous long-range spin order at 
any nonzero temperature and at any order in $t/U$ -- 
despite the three-dimensional lattice structure.
Introduction of spin-orbit coupling does allow spin ordering, 
but even then the excitation spectrum is gapless due to 
a subtle continuous symmetry. Finally we showed that these
hidden symmetries dramatically simplify the numerical
exact diagonalization studies of finite clusters.

\end{abstract}

%


\maketitle

\section{INTRODUCTION}

The transition metal oxides have been the source of many fascinating
physical phenomena such as high temperature superconductivity\cite{B},
colossal magnetoresistance\cite{CMR}, half metallic perovskites\cite{half}
and orbiton physics\cite{KKrev, TN,ORA}.
These surprising and diverse physical properties arise
from strong correlation effects in the $3d$ bands. 
As a first-step towards a better understanding of these systems,
there is much recent interest in the
magnetic properties of transition metal oxides with cubic
ABO$_{3}$ structure, where three 
$d$-orbitals are degenerate\cite{KKrev,TN} (see inset to Fig.~1). 
In cubic oxide perovskites, the
crystal field of the surrounding oxygen octahedra splits the
$d$-orbitals into a two-fold degenerate $e_{g}$ and a three-fold
degenerate $t_{2g}$ manifold. In most cases, these degeneracies
are further lifted by a cooperative Jahn-Teller (JT)
distortion\cite{KKrev}, and the low energy physics is well
described by an effective superexchange spin-only model
\cite{ANDERSON,TY1,TY2}. However, some cubic perovskites, such as
LaTiO$_3$, do
not undergo a significant JT distortion, in spite of the orbital
degeneracy\cite{noJT}.
We will mainly consider the simplest ``idealized'' version of
this model, in which the magnetic ions occupy sites whose local
site symmetry is at least tetragonal.
In these systems, the effective
superexchange model must deal with not only the spin degrees of
freedom but also the degenerate orbital degrees of
freedom\cite{KKrev,TN,KK}. The large degeneracy of the resulting
ground states may then yield rich phase diagrams, with exotic
types of order, involving a strong interplay between the spin and
orbital sectors\cite{TN,LTO,YTO}.

In the titanates, there is one $d$ electron in the $t_{2g}$
degenerate manifold, which contains the wavefunctions $|x \rangle
\equiv d_{yz}$, $|y \rangle \equiv d_{xz}$, and $|z \rangle \equiv
d_{xy}$. We will refer to $|x\rangle$, $|y \rangle$, and $|z\rangle$
as ``flavors.'' Following Kugel and Khomskii (KK) \cite{KK}, one starts
from a Hubbard model with on-site Coulomb energy $U$ and
nearest-neighbor (nn) hopping energy $t$.
For large $U$, this
model can be reduced to an effective superexchange model, which
involves only nn spin and orbital coupling, with energies of order
$\epsilon=t^2/U$. This low energy model has been the basis for
theoretical studies of the titanates.\cite{ORA,KK,GK1,GK2}
In particular, it has been
suggested \cite{GK1} that the KK Hamiltonian gives rise to an
ordered isotropic spin phase, and that an energy gap in the spin
excitations can be caused by spin-orbit interactions \cite{GK2}.
However, these papers are based on assumptions and approximations
which are hard to assess. Recently\cite{PRL} (this will be referred
to as I) we have presented rigorous arguments which show 
several unusual symmetries of the KK Hamiltonian.  Perhaps the most
striking symmetry is the rotational invariance of the total spin of
the electrons in $\alpha$-flavor orbitals
in each plane perpendicular to the $\alpha$-axis.  As reported
previously in I, this symmetry implies that the system does not
support long-range spin order at any nonzero temperature, {\it
despite the underlying three-dimensional lattice structure}.
Inclusion of spin-orbit interaction destroys the independent
rotational invariance of the spin associated with each orbital. Although
long-range order at nonzero temperature occurs when this perturbation is
included, the spin system still possesses enough symmetry that
the excitation spectrum remains gapless.\cite{TY1,TY2,BS} 
(This conclusion might be surprising, because once spin-orbit
interactions are included,
the system is expected to distinguish directions relative to those
defined by the lattice.)  
The purpose of the present paper is to extend the theory
for the KK model presented in I to the underlying three-band 
$t_{2g}$ Hubbard model, and to elaborate further on the
consequences of these results.

Briefly this paper is organized as follows.  In Sec. II we
introduce the Hubbard model, which forms the starting point
for ``idealized'' treatments of these systems, and we discuss the
KK Hamiltonian which is its byproduct.  In Sec. III we
derive the symmetry results both on the original Hubbard model
and also on the KK Hamiltonian, which are central to most of
our arguments.  In Sec. IV we give the details of the Mermin-Wagner
construction, which enables us to rigorously conclude that
this model does not support long-range order. 
In Sec.
V we show how these unusual symmetries lead to a dramatic
simplification in the numerical determination of the ground
state of a cube of eight sites governed by the KK Hamiltonian.  In
Sec.  VI we give a symmetry analysis which shows that even in the
presence of spin-orbit interactions the excitation spectrum
must have a gapless Goldstone mode. In Sec. VII we summarize our work.
Our major conclusion is that to understand real physical systems,
such as LaTiO$_3$, which do show long-range spin order and which do
have a gap in their elementary excitation spectrum, it is crucial
to include perturbations which destroy the symmetries described here.

\section{THE HAMILTONIAN}

The system we treat is a simple cubic lattice of ions with one
electron per ion in a $d$ band.  Following the seminal work of Kugel
and Khomskii\cite{KK} (KK), we describe this system by a Hubbard
Hamiltonian ${\cal H}_{\rm H}$ of the form
\begin{eqnarray}
{\cal H}_{\rm H} &=& \sum_{i \alpha} {\tilde \epsilon}_\alpha N_\alpha(i) 
+ \sum_{ij} \sum_{\alpha \beta \sigma} t_{\alpha \beta} (i,j)
c_{i \alpha \sigma}^\dagger c_{j \beta \sigma}
+ \frac 12 \sum_i \sum_{\alpha \beta} U_{\alpha \beta}
N_\alpha(i) N_\beta (i) \ ,
\label{HUBBARD} \end{eqnarray}
where $c_{i \alpha \sigma}^\dagger$ 
creates an electron in the orbital
labeled $\alpha$ in spin state $\sigma$ on site $i$ and
$N_\alpha (i) = \sum_\sigma c_{i \alpha \sigma}^\dagger c_{i \alpha \sigma}$,
${\tilde \epsilon}_\alpha \equiv \epsilon_\alpha - (U_{\alpha \alpha}/2)$,
where $\epsilon_\alpha$ is the crystal field energy of the $\alpha$
orbital, $t_{\alpha \beta}(i,j)$ (which we assume to be real) is the matrix
element for hopping between orbital $\alpha$ of site $i$ and orbital
$\beta$ of site $j$, and $U_{\alpha \beta}$ is the direct Coulomb
interaction between electrons in orbitals $\alpha$ and $\beta$ on the same
site.  Note that this Hamiltonian does not include the somewhat smaller
Coulomb exchange terms.\cite{GK,IHM}
[We write ${\cal H}_{\rm H}$ in the above form to emphasize that apart from
the hopping term, the Hamiltonian is a function of $N_\alpha (i)$.]  We
assume that the crystal field splits the five orbital $d$ states into three
low-energy $t_{2g}$ states, $d_{yz}\equiv |x\rangle$, $d_{xz}\equiv |y \rangle$, and
$d_{xy}\equiv |z \rangle$ and that the two other $e_g$ states have high enough
energy that we can neglect their presence (Fig.~1).  
This structure is consistent with tetragonal or higher site symmetry.  
(For strictly cubic site symmetry the higher two states form a degenerate 
doublet and the lower three states a degenerate triplet.  
For lower-than-cubic symmetry the degeneracy in energy of the states 
is removed but, as long as the symmetry is tetragonal or higher, 
the wave functions are those listed.) We confine our attention 
to the wide class of materials in which hopping between magnetic ions 
is mediated by intervening oxygen ions. In that case, following KK, we 
note that the hopping matrix $t_{\alpha \beta}$ between nn's is diagonal in 
orbital indices and also $t_{\alpha \alpha}=0$ if the nn bond lies along 
the $\alpha$ axis, which has been called\cite{GK} the ``inactive'' axis for
$\alpha$ hopping.  Later on we will discuss modifications caused
by further-than-nearest-neighbor hopping via oxygen ions.  The symmetry
of the nn hopping matrix is illustrated in Fig. \ref{hop}. Note that
these symmetries only hold for tetragonal or cubic site symmetry.
They are broken by a rotation of the oxygen octahedra. 

When $t \ll U$, KK reduced the above Hubbard Hamiltonian   at lowest
order in $t/U$  to an effective Hamiltonian for the manifold of
states which remain when $U \rightarrow \infty$.  We 
will call this Hamiltonian the KK Hamiltonian, and it
will be denoted ${\cal H}_{\rm KK}$.  This Hamiltonian
can be regarded as a many-band generalization of the
Heisenberg Hamiltonian, which is obtained from the single-band
Hubbard model with one electron per site, for which the
exchange constant is $J= 4 t^2/U$.  To make the analogy with
the Heisenberg model more apparent, the KK Hamiltonian is often
written in terms of spin variables as
\begin{eqnarray}
{\cal H}_{\rm KK} &=& \epsilon \sum_{\langle ij \rangle}
J_{ij} [1 + \sigmav_i \cdot \sigmav_j]
\label{KKEQ} \end{eqnarray}
where ${\bf S}_i \equiv {1 \over 2} \sigmav_i$ is the vector spin
operator for an electron on site $i$, 
$\langle ij \rangle$ indicates that the sum is over pairs of nearest
neighbor ions, $\epsilon=t^2/U$, and the exchange ``constant'' is now
an orbital operator written in terms of spinless fermion operators as
\begin{eqnarray}
J_{ij} &=& \sum_{\alpha \not= ij } \sum_{\beta \not= ij }
a_{i \alpha}^\dagger a_{i \beta} a_{j \beta}^\dagger a_{j \alpha} \ ,
\end{eqnarray}
where $a_{i\alpha}^\dagger$ creates an electron in orbital $|\alpha\rangle$
on site $i$ (of either spin) and $\alpha \not= ij$ means that we sum
$\alpha$ over the values $x$, $y$, and $z$, except for the
value of $\alpha$ which corresponds to the coordinate direction of 
the bond $ij$, which is the inactive axis for orbital $\alpha$.

For our purposes it is more convenient to write
${\cal H}_{\rm KK} = {\cal H}_{x} + {\cal H}_{y} + {\cal H}_{z}$, where
\begin{eqnarray}
{\cal H}_\alpha = \epsilon \sum_{\langle ij \rangle \in \alpha}
\sum_{\beta , \gamma \not= \alpha} \sum_{\sigma , \eta} 
c_{i \beta  \sigma}^\dagger c_{i \gamma  \eta}
c_{j \gamma  \eta}^\dagger c_{j \beta \sigma} \ ,
\label{HKK} 
\end{eqnarray}
where $\langle ij \rangle \in \alpha$ means that the sum over pairs
of nearest neighbors is restricted to those for which ${\bf r}_{ij}$
is along an $\alpha$-axis.
It should be clear that ${\cal H}_{\rm KK}$ will inherit whatever
symmetries which are present in ${\cal H}_{\rm H}$, although additional
symmetries are to be expected.  For instance, one can regard
the elimination of states in which any site is
doubly occupied as resulting from a canonical transformation
which eliminates such states.\cite{QTS}  Such a programme
can be carried out for the single band Hubbard model, even when
it is not at half filling.\cite{HL}  The important point is that
whatever rotational symmetries we uncover in ${\cal H}_{\rm H}$
should apply also to ${\cal H}_{\rm KK}$.  On the other
hand, the conservation law that each site has a single electron
will only hold for ${\cal H}_{\rm KK}$ because it is precisely
the double occupancy sites which have been eliminated by the
canonical transformation.

\section{HIDDEN ROTATIONAL SYMMETRY}

Previously in I\cite{PRL}  we pointed out
several unusual symmetries of the effective  Hamiltonian 
${\cal H}_{\rm KK}$.  Here
we extend our analysis to the underlying Hubbard model. 
It is useful to rewrite ${\cal H}_{\rm H}$ to
display explicitly the form of the hopping matrix element:
\begin{eqnarray}
{\cal H}_{\rm H} &=& \sum_{i \alpha \sigma} {\tilde \epsilon}_\alpha
N_\alpha(i) + t {\sum_{ij}}^\prime \sum_{\alpha \not= ij} \sum_\sigma
c_{i \alpha \sigma}^\dagger c_{j \alpha \sigma}
+ \frac 12 \sum_i \sum_{\alpha \beta} U_{\alpha \beta} 
N_\alpha (i) N_\beta (i) \ ,
\label{HAEQ} \end{eqnarray}
where the prime on the sum over $i$ and $j$ limits this sum to the
case when these sites are nearest neighbors of one another.
 
First we show that the total number of electrons in $\alpha$ orbitals in 
any chosen plane of sites perpendicular to the $\alpha$-axis is conserved.
For that purpose we note that the only operator which changes the
occupancy of orbital states is the hopping term, ${\bf T}$, where
\begin{eqnarray}
{\bf T} &=& t {\sum_{ij}}^\prime \sum_{\alpha \not= ij} \sum_\sigma
c_{i \alpha \sigma}^\dagger c_{j \alpha \sigma} \ .
\end{eqnarray}
Electrons in an $\alpha$-orbital may hop only to another $\alpha$-orbital.
Thus, the total number of $\alpha$-electrons (by this we mean electrons
in $\alpha$ orbitals) is a good quantum number.\cite{IHM}  Furthermore,
hopping between $\alpha$-orbitals can only take place within the same
$\alpha$-plane, i.e. within the same plane perpendicular to the
$\alpha$-axis, which is the inactive axis for
$\alpha$-flavor electrons.  Thus, the total number of $\alpha$-electrons 
in each $\alpha$-plane is a good quantum number, so that
the operator $N_\alpha(\alpha_0)$ commutes with the Hamiltonian, where
\begin{eqnarray}
N_\alpha (\alpha_0) &\equiv& \sum_{i \in \alpha_0}
\sum_\sigma c_{i \alpha \sigma}^\dagger c_{i \alpha \sigma} \ .
\end{eqnarray}
Here the notation $i  \in \alpha_0$ indicates that $i$ is
summed over all sites in the $\alpha$-plane for which
$r_{i\alpha} = \alpha_0$, where $r_{i\alpha}$ is the 
$\alpha$ component of the position of site $i$. As mentioned, this
property must be inherited by the KK Hamiltonian, and indeed one
can show this explicitly.\cite{PRL}

We now consider the much more general symmetry induced by an arbitrary
unitary transformation among spin states {\it of a given orbital flavor}.
We consider the effect of the transformation applied to operators on site $i$,
\begin{eqnarray}
c_{i \alpha \sigma}^\dagger &=& \sum_\eta  {\bf U}_{\sigma \eta}
\tilde c_{i \alpha \eta}^\dagger 
\end{eqnarray}
and
\begin{eqnarray}
c_{i \alpha \sigma}&=& \sum_\eta  {\bf U}^*_{\sigma \eta}
\tilde c_{i \alpha \eta} = \sum_\eta  {\bf U}^\dagger_{\eta \sigma}
\tilde c_{i \alpha \eta}  \ ,
\end{eqnarray}
where ${\bf U}$ is an arbitrary two dimensional unitary matrix and
$\alpha$ is a fixed flavor the choice of which is arbitrary.  Note that
\begin{eqnarray}
N_\alpha (i) &=& \sum_\sigma c_{i \alpha \sigma}^\dagger c_{i \alpha \sigma} =
\sum_{\sigma \eta \rho} {\bf U}_{\sigma \eta} {\bf U}^\dagger_{\rho \sigma}
\tilde c_{i \alpha \eta}^\dagger \tilde c_{i \alpha \rho} \nonumber \\
&=& \sum_{\eta \rho} \left[ {\bf U}^\dagger {\bf U} \right]_{\rho \eta}
\tilde c_{i \alpha \eta}^\dagger \tilde c_{i \alpha \rho} \nonumber \\
&=& \sum_{\eta} \tilde c_{i \alpha \eta}^\dagger \tilde c_{i \alpha \eta}
= {\tilde N}_\alpha (i) \ .
\end{eqnarray}
This shows that the first and third terms of Eq. (\ref{HAEQ}) are
invariant under this type of local unitary transformation applied only
to orbital $\alpha$ on site $i$.  The second term will likewise be invariant
under an arbitrary U2 transformation, providing we transform all
operators connected by hopping in the same way.  This means that we set
\begin{eqnarray}
c_{i \beta \sigma}^\dagger &=& \sum_\eta  {\bf U}_{\sigma \eta}(i , \beta)
\tilde c_{i \beta \eta}^\dagger 
\label{cUc}
\end{eqnarray}
and
\begin{eqnarray}
c_{i \beta \sigma}&=& \sum_\eta  {\bf U}^\dagger_{\eta \sigma} (i, \beta)
\tilde c_{i \beta \eta}  \ ,
\label{CUC}
\end{eqnarray}
where ${\bf U}(i,\beta)$ is the unit ($2 \times 2$) matrix, unless both
$\beta = \alpha$ and $i$ is in the given $\alpha$-plane:
$r_{i\alpha} = \alpha_0$, in which case
${\bf U}(i,\beta) = {\bf U}$, where ${\bf U}$ is an arbitrary two
dimensional unitary matrix.  Thus we apply the transformation ${\bf U}$ to
all electrons of a given orbital flavor {\it which are in an arbitrarily
chosen plane perpendicular to the inactive axis for this flavor}.
To see what this means, we write the spin operator for $\alpha$-electrons
in the chosen plane as
\begin{eqnarray}
{\bf S}_\alpha(\alpha_0) = \sum_{r_{i \alpha} = \alpha_0} \sum_{\eta \sigma}
c_{i \alpha \sigma}^\dagger \sigmav_{\sigma \eta} c_{i \alpha \eta} 
\equiv \sum_{r_{i \alpha} = \alpha_0} {\bf S}_{i\alpha}  \  , 
\label{SVECEQ} \end{eqnarray}
where ${\bf S}_{i\alpha}$ is the vector spin operator for electrons
on site $i$ of orbital flavor $\alpha$.
Then we conclude that since ${\bf U}$ is an arbitrarily chosen unitary
matrix the above discussion shows that ${\bf S}_\alpha (\alpha_0)$,
the total spin summed over all electrons in orbital $\alpha$
which are in any arbitrarily chosen $\alpha$ plane can be rotated
arbitrarily at zero cost in energy.
In other words the total spin of this orbital flavor in an arbitrarily
chosen plane perpendicular to the inactive axis of this flavor, as well
as the $z$-component of this total spin, are also good quantum numbers.
Applying the transformation (\ref{cUc}) and (\ref{CUC}) to
the KK Hamiltonian (\ref{HKK}) we note that, as should be the case, 
this Hamiltonian is 
also invariant under the rotation of the total spin of
all electrons in orbital $\alpha$ in a given $\alpha$-plane.

We now discuss these results.  First of all, the KK Hamiltonian of
Eq. (\ref{KKEQ}) is somewhat similar to the Heisenberg Hamiltonian in
that it involves the scalar product of spin operators on different sites.
This form guarantees that the Hamiltonian is invariant under a global
rotation of spin.  We have demonstrated a much stronger symmetry, which
only holds because the constant added to $\sigmav_i \cdot \sigmav_j$
in Eq.  (\ref{KKEQ}) is unity. [Only in this case is it possible
to write Eq.  (\ref{HKK}) in the alternate form of Eq. (\ref{KKEQ}).]
Furthermore, had we allowed Coulomb exchange in the
Hubbard Hamiltonian, that is, had we allowed spin exchange in the Coulomb
interaction term of Eq. (\ref{HAEQ}), then we would have global spin
rotation (just as in the Heisenberg case), but we would not have had the
invariance against independent rotations of the spins of each flavor.
In fact, any interaction which allows an electron of one orbital flavor
to convert (via hopping or some other interaction) into a different orbital
flavor will also clearly invalidate the property that allows independent
spin rotations for different orbital flavors.
Probably the most important perturbation which allows such off-diagonal
hopping in orbital flavor is the non-collinear  $M - O - M$ 
bonding due to the rotation of the $MO_{6}$-octahedra in real transition
metal oxides. This rotation angle is $156^{\rm o}$ in LaTiO$_{3}$\cite{noJT}
and therefore it has
to be taken into account in any theory to explain the observed experimental
properties of LaTiO$_{3}$.
Similarly, this symmetry is also
destroyed by spin-orbit interactions, Coulomb exchange interactions,
hopping between second nearest neighbor oxygen ions,
and less importantly by direct exchange between nearest neighboring
Ti ions and by dipolar
interactions. 

\section{MERMIN-WAGNER CONSTRUCTION}

In I we used a Mermin-Wagner construction\cite{MW} to establish
the absence of long-range spin order due to the KK Hamiltonian in
three (or less) dimensions, but we did not give full
details of this construction.  Here we present such
a construction, but obtain a more powerful result by working
with the Hubbard Hamiltonian.  Results obtained from the KK Hamiltonian
are only valid to leading order in $t/U$.  By working with the Hubbard
model Hamiltonian, we will establish these results to all orders in $t/U$
for the Hamiltonian of Eq. (\ref{HAEQ}).

The Mermin-Wagner construction relies on the Bogoliubov inequality,\cite{BOG}
which is
\begin{eqnarray}
{1 \over 2} \langle [[C,{\cal H}]_-, C^\dagger]_- \rangle 
\langle \{ A,A^\dagger\}_+ \rangle \geq kT  |\langle [C,A]_-\rangle|^2 \ .
\label{B} \end{eqnarray}
where $\langle X \rangle$ denotes the canonical average of the operator $X$
at temperature $T$,
$[X,Y]_-$ is the commutator of $X$ and $Y$ and $\{X,Y\}_+$ is the
anticommutator of $X$ and $Y$.  As we shall see, it is also crucial
to note that
\begin{eqnarray}
C_{\cal H} \equiv \langle [[C,{\cal H}]_-, C^\dagger]_- \rangle > 0  \ ,
\label{POSITIVE} \end{eqnarray}
unless $[C,{\cal H}]_- =0$, in which $C_{\cal H}=0$.  To verify this
we write out the expression for $C_{\cal H}$ in terms of the system's exact
energy eigenstates, $|n\rangle$:
\begin{eqnarray}
C_{\cal H} &=& \sum_{n,m} p_n \Biggl[ \langle n|C|m\rangle (E_m-E_n) 
\langle m | C^\dagger | n \rangle - \langle n | C^\dagger |m \rangle
\langle m | C | n \rangle (E_n-E_m) \Biggr] \ ,
\label {CSUBH} \end{eqnarray}
with $p_n = e^{-\beta E_n}/Z$, where $Z$ is the partition function,
\begin{eqnarray}
Z = \sum_n e^{-\beta E_n} \ ,
\end{eqnarray}
and $\beta = 1/(kT)$. In the second term of Eq. (\ref{CSUBH})
we interchange the roles of $n$ and $m$, after which we obtain
\begin{eqnarray}
C_{\cal H} &=& \sum_{n,m} |\langle n|C|m\rangle|^2
[ p_n - p_m][E_m-E_n] \ .
\label{SUMMAND} \end{eqnarray}
We have that
\begin{eqnarray}
[p_n-p_m][E_m-E_n] &=& Z^{-1} e^{-\beta (E_n+E_m)/2}
\left( 2 E_{n,m}\sinh[(\beta E_{n,m}/2] \right) \ ,
\end{eqnarray}
where $E_{n,m}=E_n-E_m$.  Thus the summand in Eq. (\ref{SUMMAND}) is
nonnegative.  In fact, for $C_{\cal H}$ to be zero, one has to have
$\langle n|C|m\rangle=0$ whenever $E_n \not=E_m$.  In other words,
$C_{\cal H}$ is real and nonnegative and can only be zero if
the operator $C$ commutes with the Hamiltonian, a case we
will not encounter here.

For the purposes of this construction we will add to the Hamiltonian
a term conjugate to the magnetization at wavevector ${\bf K}$
of electrons in $\alpha$-orbitals, so that we treat the Hamiltonian
\begin{eqnarray}
{\cal H} &=& {\cal H}_{\rm H} - {1 \over 2}
h \sum_{\bf R} e^{i {\bf K} \cdot {\bf R}}
[c_{{\bf R} \alpha \uparrow}^\dagger c_{{\bf R} \alpha \uparrow}
- c_{{\bf R} \alpha \downarrow}^\dagger c_{{\bf R} \alpha \downarrow}]
\equiv {\cal H}_{\rm H} - h \sum_{\bf R} e^{i {\bf K} \cdot {\bf R}}
S_{{\bf R} \alpha}^z  \nonumber \\
&\equiv & {\cal H}_{\rm H} - h N S_{{\bf K} \alpha}^z  \ ,
\end{eqnarray}
where ${\bf S}_{{\bf R}\alpha}$ is the spin operator for the $\alpha$ orbital
of site ${\bf R}$, as in Eq. (\ref{SVECEQ}).  We will set the order-parameter
wavevector ${\bf K}$ to be zero to exclude ferromagnetic long-range order and
to be $(\pi , \pi, \pi)/a$ to exclude antiferromagnetic
long-range order, where $a$ is the lattice constant. 
However, as can be seen later, we may choose ${\bf K}$
arbitrarily, in which case our construction excludes the
possibility of long-range order at whatever ${\bf K}$ is chosen.

To rule out long-range spin ordering, we apply Eq. (\ref{B}) taking
\begin{eqnarray}
C_{\bf k} &=& \sum_{\bf R} e^{-i {\bf k} \cdot {\bf R} }
c_{{\bf R} \alpha \uparrow}^\dagger c_{{\bf R} \alpha \downarrow}\ ,
\end{eqnarray}
\begin{eqnarray}
A_{\bf k} &=& \sum_{\bf R} e^{i ({\bf k} + {\bf K}) \cdot {\bf R}}
c_{{\bf R} \alpha \downarrow}^\dagger c_{{\bf R} \alpha \uparrow}\ ,
\end{eqnarray}
where ${\bf R}$ is a lattice site.  Note that the operator
$c_{{\bf R}\alpha \uparrow}^\dagger c_{{\bf R}\alpha \downarrow}$ can be
identified as $S_{{\bf R}\alpha}^+$, the raising operator for
${\bf S}_{{\bf R}\alpha}$.  Then Eq. (\ref{B}) gives
\begin{eqnarray}
\sum_{\bf k} \langle \{ A_{\bf k},A_{\bf k}^\dagger\}_+ \rangle \geq
2kT \sum_{\bf k} { |\langle [C_{\bf k},A_{\bf k}]_-\rangle|^2 \over
\langle [[C_{\bf k},{\cal H}]_-, C_{\bf k}^\dagger]_- \rangle } \ .
\end{eqnarray}
Here and below the sum over ${\bf k}$ is over the $N$ wavevectors of the
first Brillouin zone, where $N$ is the total number of sites.  We have
\begin{eqnarray}
\sum_{\bf k} \langle \{ A_{\bf k},A_{\bf k}^\dagger\}_+ \rangle 
= \sum_{\bf k} \sum_{\bf R} \sum_{\bf S}
e^{i ({\bf k } + {\bf K} ) \cdot {\bf R}}
e^{-i ({\bf k } + {\bf K} ) \cdot {\bf S}}
\langle \{ c_{{\bf R} \alpha \downarrow}^\dagger
c_{{\bf R} \alpha \uparrow} , c_{{\bf S} \alpha \uparrow}^\dagger
c_{{\bf S} \alpha \downarrow} \}_+ \rangle \ .
\end{eqnarray}
The sum over ${\bf k}$ gives $N \delta_{{\bf R}, {\bf S}}$, so that
\begin{eqnarray}
\sum_{\bf k} \langle \{ A_{\bf k},A_{\bf k}^\dagger\}_+ \rangle 
= N \sum_{\bf R} \langle \{ c_{{\bf R} \alpha \downarrow}^\dagger
c_{{\bf R} \alpha \uparrow} , c_{{\bf R} \alpha \uparrow}^\dagger
c_{{\bf R} \alpha \downarrow} \}_+ \rangle \ .
\end{eqnarray}
Here the sum over ${\bf R}$ consists of a sum of $2N$ products, each of
which is bounded by unity.  Thus we have
\begin{eqnarray}
2N^2 \geq \sum_{\bf k} \langle \{ A_{\bf k},A_{\bf k}^\dagger\}_+ \rangle 
\end{eqnarray}
which gives 
\begin{eqnarray}
N^2 \geq  kT \sum_{\bf k} { |\langle [C_{\bf k},A_{\bf k}]_-\rangle|^2
\over \langle [[C_{\bf k},{\cal H}]_-, C_{\bf k}^\dagger]_- \rangle } \ .
\end{eqnarray}

Now we evaluate 
\begin{eqnarray}
[C_{\bf k},A_{\bf k}]_- &=&
\sum_{{\bf R}, {\bf S}} e^{-i {\bf k} \cdot {\bf R} }
e^{i ({\bf k} + {\bf K}) \cdot {\bf S}}
[c_{{\bf R} \alpha \uparrow}^\dagger c_{{\bf R} \alpha \downarrow} ,
c_{{\bf S} \alpha \downarrow}^\dagger
c_{{\bf S} \alpha \uparrow} ]_- \nonumber \\ &=&
\sum_{\bf R} e^{i {\bf K} \cdot {\bf R} }
[c_{{\bf R} \alpha \uparrow}^\dagger c_{{\bf R} \alpha \downarrow} ,
c_{{\bf R} \alpha \downarrow}^\dagger c_{{\bf R} \alpha \uparrow} ]_- \ . 
\end{eqnarray}
Using
\begin{eqnarray}
[ AB, CD]_- &=& A \{B,C\}_+D - \{A,C \}_+ BD + CA \{B,D\}_+ - C\{A,D\}_+B \ .
\label{COMM} \end{eqnarray}
we then get
\begin{eqnarray}
[c_{{\bf R} \alpha \uparrow}^\dagger c_{{\bf R} \alpha \downarrow} ,
c_{{\bf R} \alpha \downarrow}^\dagger c_{{\bf R} \alpha \uparrow} ]_-
&=& c_{{\bf R} \alpha \uparrow}^\dagger c_{{\bf R} \alpha \uparrow}
- c_{{\bf R} \alpha \downarrow}^\dagger c_{{\bf R} \alpha \downarrow}
\end{eqnarray}
(which is nothing more than $[S^+_{{\bf R}\alpha}, S^-_{{\bf R}\alpha}]_-
=2 S^z_{{\bf R}\alpha}$), so that
\begin{eqnarray}
\langle [C_{\bf k},A_{\bf k}]_- \rangle &=&
2 \sum_{\bf R} e^{i {\bf K} \cdot {\bf R} } s_z (\alpha, {\bf R})
= 2 N S_{{\bf K} \alpha}^z \ .
\end{eqnarray}
For $K=0$ $S_{{\bf K} \alpha}^z$ is the $z$-component of the spin per site
of $\alpha$-flavor orbitals and for ${\bf K} = (\pi , \pi , \pi)a$ it is
the $z$-component of the staggered magnetization per site of $\alpha$-flavor
orbitals.  (For general ${\bf K}$ it is the amplitude of the Fourier
component of the $z$-component of spin of $\alpha$-flavor orbitals at
wavevector ${\bf K}$.)  We concentrate on the ferromagnetic and antiferromagnetic
cases.  We have the inequality
\begin{eqnarray}
1\geq  4kT |S_{{\bf K}\alpha}^z|^2 \sum_{\bf k} \langle X_{\bf k} \rangle ^{-1} \ ,
\label{INEQUAL} \end{eqnarray}
where $X_{\bf k}$ is positive real and is given by
\begin{eqnarray}
X_{\bf k} &=& [[C_{\bf k},{\cal H}]_-, C_{\bf k}^\dagger]_- \ .
\label{COMMUTATE} \end{eqnarray}

Now we need the double commutator of $C_{\bf k}$ with ${\cal H}$.
The Hamiltonian is the sum of four terms: $T_1$ involving the
hopping matrix element $t_{ij}^\alpha$, $T_2$ involving the
crystal field energies $\epsilon_\alpha$, $T_3$ the on-site
Coulomb interaction, and $T_4$ the interaction with the
staggered field $h>0$. Since $T_2$ and $T_3$ depend only on
$N_\alpha (i)$, one can establish that
\begin{eqnarray}
[C_{\bf k}, T_2]_- = [C_{\bf k}, T_3]_- = 0 \ .
\end{eqnarray}
Also
\begin{eqnarray}
[C_{\bf k}, T_4]_- &=& h \sum_{\bf R} e^{i ( {\bf K} - {\bf k})
\cdot {\bf R}} c_{{\bf R} \alpha \uparrow}^\dagger
c_{{\bf R} \alpha \downarrow} 
\end{eqnarray}
so that
\begin{eqnarray}
\langle [[C_{\bf k}, T_4]_-,C_{\bf k}^\dagger]_- \rangle &=&
\langle h \sum_{\bf R} e^{i {\bf K} \cdot {\bf R}}
[c_{{\bf R} \alpha \uparrow}^\dagger c_{{\bf R} \alpha \downarrow} \ ,
c_{{\bf R} \alpha \downarrow}^\dagger c_{{\bf R} \alpha \uparrow}]_-
\rangle \nonumber \\
&=& h \sum_{\bf R} e^{i {\bf K} \cdot {\bf R}} \langle [
c_{{\bf R} \alpha \uparrow}^\dagger c_{{\bf R} \alpha \uparrow}
- c_{{\bf R} \alpha \downarrow}^\dagger c_{{\bf R} \alpha \downarrow} ]
\rangle \nonumber \\ &=& 2Nh S_{{\bf K}\alpha}^z \ .
\end{eqnarray}

The crucial calculation is the commutator involving $T_1$:
\begin{eqnarray}
Y_{\bf k} \equiv [ C_{\bf k},T_1]_- = \sum_{\bf R} e^{-i {\bf k} \cdot {\bf R}}
[ c_{{\bf R} \alpha \uparrow}^\dagger c_{{\bf R} \alpha \downarrow}
, \sum_{{\bf S} , {\bf T}, \sigma} t_\alpha ({\bf S},{\bf T})
c_{{\bf S} \alpha \sigma}^\dagger c_{{\bf T} \alpha \sigma} ]_- \ ,
\end{eqnarray}
where ${\bf R}$, ${\bf S}$, and ${\bf T}$ are lattice sites
and we dropped terms involving orbitals other than $\alpha$
since they obviously commute with $C_{\bf k}$. Also in this
section we set $t_\alpha({\bf R}, {\bf S})=t_\alpha({\bf R}-{\bf S})$
to denote the hopping matrix element (assumed to be nonnegative for
convenience) between $\alpha$ orbitals
on sites ${\bf R}$ and ${\bf S}$ and we use $t_\alpha({\bf R})
= t_\alpha(-{\bf R})$.

We have
\begin{eqnarray}
Y_{\bf k} &=& \sum_{{\bf R} {\bf S} {\bf T} \sigma}
e^{-i {\bf k} \cdot {\bf R}} t_\alpha ({\bf S}, {\bf T})
\left[ c_{{\bf R} \alpha \uparrow}^\dagger c_{{\bf R} \alpha \downarrow} ,
c_{{\bf S} \alpha \sigma}^\dagger c_{{\bf T} \alpha \sigma} \right]_-
\nonumber \\ &=&  \sum_{{\bf R} {\bf S} {\bf T} } 
e^{-i {\bf k} \cdot {\bf R}} t_\alpha ({\bf S}, {\bf T})
\left[ c_{{\bf R} \alpha \uparrow}^\dagger \delta_{{\bf R}, {\bf S}}
c_{{\bf T} \alpha \downarrow} -
c_{{\bf S} \alpha \uparrow}^\dagger \delta_{{\bf R}, {\bf T}}
c_{{\bf R} \alpha \downarrow} \right] \nonumber \\
&=& \sum_{{\bf R} \deltav} e^{-i {\bf k} \cdot {\bf R}}
t_\alpha(\deltav) \left[ 1 - e^{- i {\bf k} \cdot \deltav} \right]
c_{{\bf R} \alpha \uparrow}^\dagger c_{{\bf R} + \deltav \alpha \downarrow} \ ,
\end{eqnarray}
where $\deltav$ is a nearest-neighbor vector.  Then
\begin{eqnarray}
X_{\bf k} &\equiv & [[C_{\bf k}, T_1]_- , C_{\bf k}^\dagger]_- +2Nh S_{{\bf K}\alpha}^z
= [ Y_{\bf k} , C_{\bf k}^\dagger ]_-  +2Nh S_{{\bf K}\alpha}^z
\nonumber \\
&=& \sum_{{\bf R} {\bf S} \deltav} e^{i {\bf k} \cdot ({\bf S}- {\bf R})}
t_\alpha (\deltav) \left[ 1 - e^{-i {\bf k} \cdot \deltav} \right]
\left[ c_{{\bf R} \alpha \uparrow}^\dagger
c_{{\bf R}+\deltav \alpha \downarrow} ,
c_{{\bf S} \alpha \downarrow}^\dagger
c_{{\bf S} \alpha \uparrow} \right]_-
+ 2Nh S_{{\bf K}\alpha}^z
\nonumber \\ &=& \sum_{{\bf R} \deltav}
t_\alpha (\deltav) \left[ 1 - e^{-i {\bf k} \cdot \deltav} \right]
\left[ e^{i {\bf k} \cdot \deltav} c_{{\bf R} \alpha \uparrow}^\dagger
c_{{\bf R}+\deltav \alpha \uparrow} -
c_{{\bf R} \alpha \downarrow}^\dagger
c_{{\bf R}+\deltav \alpha \downarrow} \right] + 2Nh S_{{\bf K}\alpha}^z
\nonumber \\ &=& \sum_{{\bf R} \deltav}
t_\alpha (\deltav) \Biggl[ \left( e^{i {\bf k} \cdot \deltav} - 1 \right)
c_{{\bf R} \alpha \uparrow}^\dagger c_{{\bf R}+\deltav \alpha \uparrow} 
+ \left( e^{-i {\bf k} \cdot \deltav} - 1 \right)
c_{{\bf R} \alpha \downarrow}^\dagger
c_{{\bf R}+\deltav \alpha \downarrow} \Biggr] + 2Nh S_{{\bf S}\alpha}^z \ .
\label{PARITY} \end{eqnarray}

According to Eq. (\ref{POSITIVE}) the quantity $\langle X_{\bf k} \rangle$
is real and positive (no matter what the value of ${\bf k}$)
and therefore we can maintain the
inequality (\ref{INEQUAL}) if we replace the sum $\cal{S}$, where
\begin{eqnarray}
\cal{S} &\equiv & \sum_{\bf k} {1 \over \langle X_{\bf k} \rangle } \ ,
\end{eqnarray}
by a quantity which is less than $\cal{S}$.  To do that we write
\begin{eqnarray}
\cal{S} &\equiv & \sum_{\bf k}{1 \over 2}\Biggl(
{1 \over \langle X_{\bf k} \rangle } +
{1 \over \langle X_{-{\bf k}} \rangle } \Biggr)
\nonumber \\ & \geq & \sum_{\bf k}
{2 \over \langle X_{\bf k} \rangle + \langle X_{-{\bf k}} \rangle } \ .
\end{eqnarray}

This follows from the fact that when
$a$ and $b$ are positive, then $a^{-1} + b^{-1} \geq 4/(a+b)$.
Thus
\begin{eqnarray}
1\geq  8kT |S_{{\bf K}\alpha}^z|^2 \sum_{\bf k} \left[
\langle X_{\bf k} \rangle + \langle X_{-{\bf k}} \rangle \right]^{-1} \ .
\label{INEQUALB} \end{eqnarray}
Here
\begin{eqnarray}
\left[ \langle X_{\bf k} \rangle + \langle X_{-{\bf k}} \rangle \right]/2
&=& \sum_{{\bf R} \deltav} t_\alpha (\deltav) \Biggl[
[ \cos( {\bf k} \cdot \deltav )-1] \left(
\langle c_{{\bf R}\alpha \uparrow}^\dagger
c_{{\bf R}+ \deltav \alpha \uparrow} \rangle
+ \langle c_{{\bf R} \alpha \downarrow}^\dagger
c_{{\bf R}+ \deltav  \alpha \downarrow} \rangle \right) \Biggr] \nonumber \\ &&
\ + 2N h S_{{\bf K}\alpha}^z \ .
\label{XPXM} \end{eqnarray}

Since the quantity on the right-hand side of Eq. (\ref{XPXM})
is positive, we may replace it (without increasing $\cal{S}$) by the sum of
the absolute values
of bounds to its terms, so that finally Eq. (\ref{INEQUALB}) yields
\begin{eqnarray}
4kT |S_{{\bf K}\alpha}^z |^2 \leq \Biggl[ {1 \over N} \sum_{\bf k}
\left( \sum_\deltav t_\alpha (\deltav )
\left[ 1 - \cos({\bf k} \cdot \deltav ) \right] + h |S_{{\bf K}\alpha}^z |
\right)^{-1} \Biggr]^{-1}  = t/I \ ,
\end{eqnarray}
where
\begin{eqnarray}
I &=& {t \over N} \sum_{\bf k} \Biggl( \sum_\deltav t_\alpha (\deltav)
[1 - \cos ({\bf k} \cdot \deltav) ] + h |S_{{\bf K}\alpha}^z | \Biggr)^{-1} \ ,
\label{BOUND} \end{eqnarray}
and we have used the bound
\begin{eqnarray}
|\langle c_{{\bf R}+\deltav \alpha \uparrow}^\dagger
c_{{\bf R} \alpha \uparrow} \rangle
+ \langle c_{{\bf R} + \deltav \alpha \downarrow}^\dagger
c_{{\bf R} \alpha \downarrow} \rangle| \leq 2C_0 \ ,
\label{CRUDE} \end{eqnarray}
where $C_0=1$.  In I we gave the result obtained in an entirely analogous
fashion from the KK Hamiltonian.  That result is valid for the KK
Hamiltonian, which itself assumes the validity of the expansion in
powers of $t/U$.  That result could be obtained here by noting
that to lowest order in perturbation theory in $t/U$, the left-hand
side of Eq.  (\ref{CRUDE}) is given by\cite{PERT} $C_0 \sim 2|t|/U$.  The present
result avoids any rigorous discussion of the validity of the
expansion in powers of $t/U$, but uses the much cruder bound
of Eq. (\ref{CRUDE}) with $C_0=1$. In any event, an entirely
analogous construction can be carried out for the KK Hamiltonian,
as discussed in I.

The analysis can now be less formal. We estimate the quantity $I$
on the right-hand side of Eq. (\ref{BOUND}).  For $\alpha=z$,
for example, we have
\begin{eqnarray}
\sum_\deltav t_\alpha (\deltav) [ 1 - \cos ({\bf k} \cdot \deltav] &=&
2t [ 2 - \cos (k_xa) - \cos (k_ya) ]\ . 
\end{eqnarray}
Since the sum over $\bf k$ is dominated by small $|{\bf k}|$, we may write
\begin{eqnarray}
I & = &  A \int_0^{x_0} {2k_\perp d k_\perp \over
k_\perp^2 +  Bh|S_{{\bf K}z}^z|/t }  \nonumber \\ &=&
A \ln \Biggl[ 1 + {tx_0^2 \over Bh|S_{{\bf K}z}^z|} \Biggr]\ ,
\end{eqnarray}
where $A$,$B$, and $x_0$ are constants of order unity, whose exact
values need not concern us.  In all we obtain the bound
\begin{eqnarray}
|S_{{\bf K}z}^z| \leq A'T^{-1/2} | \ln |h|/t|^{-1/2} \ ,
\end{eqnarray}
where $A'$ is a constant.  Obviously, this bound applies also to any
of the other components, $S_{{\bf K}\alpha}^\mu$, of the
spin of any flavor $\alpha$.
This bound implies that as $h \rightarrow 0$, we must
have $|S_{{\bf K}\alpha}^\mu| \rightarrow 0$, for all choices
of $\bf K$, $\alpha$, and $\mu$.  Thus we conclude that
this Hubbard model (and perforce also the KK Hamiltonian obtained
from it) can not support spontaneous long-range spin order at any
nonzero temperature.
Our rigorous arguments have nothing to say about orbital order.
However, we believe that there is no spontaneous long-range
orbital order for the KK model with no spin-orbit interactions.

From the form of the bound we also expect
that the lower critical dimension for the appearance of long-range
spin order is $d_<=3$. As will be discussed elsewhere,\cite{AA2}
this conclusion can also be understood within a renormalization
group analysis.

The above development allows us to make some comments on whether or
not there can be spontaneous breaking of parity symmetry.
Note that the average of the quantity on the right-hand side of Eq. (\ref{PARITY})
must be positive. However, apart from the term proportional to $h$,
this quantity vanishes for ${\bf k}=0$ and has a term linear
in ${\bf k}$.  The term linear in ${\bf k}$ gives a contribution
to $\langle X_{\bf k} \rangle$ of
\begin{eqnarray}
\delta \langle X_{\bf k} \rangle &=&
i \sum_{{\bf R} \deltav}  t_\alpha (\deltav) ({\bf k} \cdot \deltav)
\langle c_{{\bf R} \alpha \uparrow}^\dagger
c_{{\bf R}+\deltav \alpha \uparrow} -
c_{{\bf R} \alpha \downarrow}^\dagger
c_{{\bf R}+\deltav \alpha \downarrow} \rangle \ .
\label{COND} \end{eqnarray}
This quantity must be real, but must also go to zero as $h\rightarrow 0$,
in order that $\langle X_{\bf k}\rangle$ always be positive.
If parity is preserved, then, of course, 
$\langle c_{{\bf R} \alpha \sigma}^\dagger c_{{\bf R}+\deltav \alpha
\sigma} \rangle = \langle c_{{\bf R} \alpha \sigma}^\dagger
c_{{\bf R}-\deltav \alpha \sigma} \rangle$ and the term linear in ${\bf k}$
does vanish.  The converse, is not quite proven, because the
quantity in Eq. (\ref{COND}) could vanish without parity
being maintained.  So, this development rules out a
spin-independent breaking of parity.   In addition, we remind the
reader that we can choose ${\bf K}$ arbitrarily, after which
the above argument rules out spontaneous helical spin order.

\section{EXACT DIAGONALIZATION}

In this section we show that the hidden symmetries discussed above
are very useful in simplifying the exact numerical studies of small
clusters.  Such cluster studies are very important, because they
provide a way of checking analytical results and also give insight
into the nature of the ground state when there is no long range order.
The main problem in such studies of finite clusters is usually the
large matrix sizes that have to be diagonalized.  For example, to
treat the KK Hamiltonian for a cube of eight sites requires the
diagonalization of a matrix of dimensionality $6^8 \approx 1.7$ million.
Since the KK Hamiltonian is rotationally invariant, one can obtain
the full spectrum by working within the subspace of
$\sum_{i=1}^{8} S_{i}^{z}=0$, where the size of the Hamiltonian matrix
is now reduced to $70\times 3^8 \approx 1/2$ million.  The spin
degeneracy of each individual state in this manifold is $2 S + 1$,
where $S(S+1)$ is the square of the total spin of the wavefunction.

Initially we obtained the low-energy spectrum of eigenstates for a cube
of eight Ti ions by diagonalizing the 1/2 million dimensional matrix described above.
Since the Hamiltonian matrix is very sparse, it is possible to obtain the
eigenvalues and the corresponding eigenvectors in a small interval
of the spectrum starting from the lowest eigenvalue, using a standard
sparse matrix diagonalization routine. The ground state is found to be
three-fold degenerate, with energy $-6.2716 \epsilon$.
By analyzing the wavefunctions we found that the 
ground state had a total spin $S = 0$ and either
$N_x =4, N_y=4, N_z = 0$, or the two cyclic permutations of these quantum numbers,
where $N_\alpha$ is the total number of $\alpha$-flavor electrons.

The same energy spectrum and the corresponding wavefunctions can also be
found by diagonalizing much smaller matrices if one works within a manifold
defined by the conserved numbers applied to each face of the cube (see Fig.
\ref{cube}).  We actually have eighteen conserved quantum numbers
and there is an
astonishing numerical simplification  when maximal use is made of these
symmetries. For example,   the ground state found by dealing with the
matrix of dimensionality 1/2 million could alternatively be found by
diagonalizing the Hamiltonian matrix within a manifold of just 16 states!
Given the fact that the ground state has $S=0$ and $N_x=N_y=4$, one can
easily conclude that this corresponds to having $N_{x,1}=2$ electrons
in $x$ orbitals in the first $x$-plane and $N_{x,2}=2$ electrons in 
$x$ orbitals in the second $x$-plane.  For any asymmetric choice, such as
$N_{x,1}=1$ and $N_{x,2}=3$, for instance, one would have an additional
degeneracy associated with interchanging $N_{x,1}$ and $N_{x,2}$.
Thus we were sure that these states found numerically had to have the
quantum numbers $N_{x,1}=N_{x,2}=N_{y,1}=N_{y,2}=2$.  In addition,
the total spin of the two $x$ electrons in each $x$-plane had to be
zero in order to be consistent with the lack of spin degeneracy.
Thus it was clear that the ground state had to consist of a sum of
terms, each term being a product of four spin singlets.  We define
\begin{eqnarray}
|(ij)_x \rangle \equiv 2^{-1/2} [ c_{ix\uparrow}^\dagger c_{jx\downarrow}^\dagger
- c_{ix\downarrow}^\dagger c_{jx\uparrow}^\dagger ] | {\rm vac} \rangle =
|(ji)_x \rangle \ ,
\end{eqnarray}
with a similar definition of $(ij)_y$, where $|{\rm vac}\rangle$ is the
vacuum state.  Then the ground state consists of a linear combination of
terms, each term being a product of four dimer singlets which we write as
\begin{eqnarray}
|(ij)_x (kl)_x (mn)_y (op)_y \rangle \ ,
\end{eqnarray}
where the sites $i$ and $j$ are in the first $x$-plane,
$k$ and $l$ are in the second $x$-plane,
$m$ and $n$ are in the first $y$-plane,
and $o$ and $p$ are in the second $y$-plane.
The dominant configurations in the ground state are the dimer states with
the lowest expectation value of the Hamiltonian, namely the states in panels
(b) and (c) of Fig. \ref{cube}.  From one of these, say, the one shown in
panel (c), one can generate the manifold of states which can be obtained
by hopping along one or more of the $z$-directed bonds, labeled $a$, $b$,
$c$, and $d$ in the figure.  The Hamiltonian matrix within this manifold of 
16 states, is given explicitly in Table I.\cite{CUBE}  To construct this matrix
we used the fact that
\begin{eqnarray}
\langle (ij)_\alpha | {\cal H}_{\rm KK} | (ij)_\alpha \rangle 
\end{eqnarray}
is $-\epsilon$ if sites $i$ and $j$ are nn's along an axis
which is {\it not} inactive for $\alpha$ orbitals. Otherwise
this matrix element is zero.  (For instance, diagonal dimers
like the $X$ and $Y$ dimers involved in the hopping shown in
panel (d) of Fig. \ref{cube} make zero contribution to the diagonal
matrix element.  So configuration (d) has a diagonal matrix
element of $-2\epsilon$ due to the nn X and Y dimers.)
Also the matrix element
\begin{eqnarray}
\langle (ij)_x (mn)_y |  {\cal H} | (mj)_x (in)_y \rangle 
\end{eqnarray}
has the value $-\epsilon$ if the sites $i$ and $m$ which are involved in the hopping
are nn's along the active ($z$) direction, and is zero otherwise.  It is amazing to us that the
original Hamiltonian, which was obtained from a matrix of dimensionality
approximately 1/2 million, could be reduced by symmetry considerations to
an eigenvalue problem of dimension 16.  The existence of these symmetries
was numerically confirmed when we found the same value for the ground
state energy from the 16 dimensional matrix as from the full matrix.
The ground-state wavefunction,
in terms of the states in the order used for the matrix, is
\begin{eqnarray}
( \alpha, \beta , \beta,\beta, \beta, \gamma , \gamma ,\delta ,\delta,
\gamma , \gamma, \beta , \beta , \beta, \beta , \alpha) \ ,
\end{eqnarray}
where $\alpha=0.40915$, $\beta=0.23235$, $\gamma=0.21758$, and $\delta=0.14819$.

\section{SYMMETRY, SPIN-ORBIT INTERACTION, AND THE EXCITATION SPECTRUM}

Here we consider the addition of a perturbation which breaks
the symmetry whereby each orbital flavor of spin can independently
be rotated at zero cost in energy.  For concreteness we consider
the spin-orbit interaction.  Possibly one's first intuition about
the effect of spin-orbit interaction would be that it would cause
the system to exhibit long-range order (this is correct) and that
the elementary excitation spectrum would have a gap, because
the orbits would define a set of favored directions.  In
a study of a similar Hubbard model for the cuprates this second
conclusion was shown to be false.\cite{TY1}  In fact, there
it was shown that addition of spin-orbit interactions to
an isotropic Hubbard model did {\it not} lead to a gap in
the spin-wave spectrum, but that a gap does result with the
addition of both spin-orbit and Coulomb exchange interactions
(these are sometimes called Hund's rule coupling).  Here
we will establish the first part of this scenario, namely
that adding only spin-orbit  interactions to the KK
Hamiltonian or to the Hubbard model of Eq. (\ref{HUBBARD})
does {\it not} lead to a gap in the excitation spectrum.

Following Ref. \onlinecite{TY1} we now introduce a transformation
to pseudo-spin which yields a rotationally invariant Hamiltonian.
We write\cite{ERROR}
\begin{eqnarray}
c_{i \alpha \sigma} = \sum_\eta [\sigmav_\alpha]_{\sigma \eta}
d_{i \alpha \eta} \ ,
\end{eqnarray}
where $d_{i \alpha \eta}^\dagger$ creates an electron in orbital
$\alpha$ of site $i$ with {\it pseudospin} $\eta$.  Since
pseudospin is not the most intuitive concept, we will here
give a discussion that avoids use of this quantity.
In terms of ``real'' spin we will show that
the Hamiltonian is invariant under the transformation
\begin{eqnarray}
c_{i \alpha \sigma} &=& \sum_\tau {\bf V}^{(\alpha)}_{\sigma \tau}
\tilde c_{i \alpha \tau} \ ,
\end{eqnarray}
where 
\begin{eqnarray}
{\bf V}^{(\alpha)} = \sigma_\alpha {\bf U} \sigma_\alpha\ ,
\label{TRANS} \end{eqnarray}
with an arbitrary unitary matrix, ${\bf U}$. 

We now consider the effect of this transformation on the Hubbard model
of Eq. (\ref{HUBBARD}).  The Coulomb interaction  and single-site
crystal-field energy are both clearly invariant
under this transformation, because they only depend on the
total number of electrons in each orbital, 
$\sum_\sigma c_{i \alpha \sigma}^\dagger c_{i \alpha \sigma}$,
and this quantity is not changed by this transformation.  Since hopping is
diagonal in the orbital indices, the hopping term is also invariant under this
transformation.  Finally, we study how the additional spin-orbit interaction
transforms.  We write this interaction as
\begin{eqnarray}
V_{\rm S-O} &=& \lambda\sum_i \sum_{\alpha \beta \gamma} \sum_{\mu \nu}
\langle \alpha | L_\gamma | \beta \rangle [\sigmav_\gamma]_{\mu \nu}
c_{i \alpha \mu}^\dagger c_{i \beta \nu} \ ,
\end{eqnarray}
where $\langle \alpha | L_\gamma | \beta\rangle$ is  the 
orbital angular momentum matrix element, and is nonzero only when
$\alpha$, $\beta$ and $\gamma$ are all different:
\begin{eqnarray}
\langle \alpha | L_\beta | \gamma \rangle = - i 
\epsilon_{\alpha \beta \gamma} \ .
\end{eqnarray}
Thus
\begin{eqnarray}
V_{\rm S-O} &=& \lambda\sum_i \sum_{\alpha \beta \gamma} \sum_{\mu \nu}
\sum_{\rho \tau} \langle \alpha | L_\gamma | \beta \rangle
[\sigmav_\gamma]_{\mu , \nu} \left[ {\bf V}^{(\alpha)}_{\mu \rho} \right]^*
\tilde c_{i \alpha \rho}^\dagger {\bf V}^{(\beta)}_{\nu \tau}
\tilde c_{i \beta \tau} \nonumber \\
& \equiv & \lambda\sum_i \sum_{\alpha \beta \gamma} \sum_{\rho \tau}
\langle \alpha | L_\gamma | \beta \rangle {\bf M}_{\rho \tau}
\tilde c_{i \alpha \rho}^\dagger \tilde c_{i \beta \tau} \ ,
\end{eqnarray}
where
\begin{eqnarray}
{\bf M}_{\rho \tau} &=& \sum_{\mu \nu}  
[\sigmav_\gamma]_{\mu , \nu} \left[ {\bf V}^{(\alpha)}\right]_{\mu \rho}^*
{\bf V}^{(\beta)}_{\nu \tau} \nonumber \\ &=&
\Biggl( \bigl[ {\bf V}^{(\alpha)} \bigr]^\dagger \bigl[\sigmav_\gamma \bigr]
\bigl[ {\bf V}^{(\beta)} \bigr] \Biggr)_{\rho \tau} \nonumber \\ &=&
\left[ \sigmav_\alpha {\bf U}^\dagger \sigmav_\alpha \sigmav_\gamma
\sigmav_\beta {\bf U} \sigmav_\beta \right]_{\rho \tau} \ .
\end{eqnarray}
The orbital matrix element $\langle \alpha | L_\gamma|\beta \rangle$
guarantees that all three of these indices are different.
In that case we set
\begin{eqnarray}
\sigmav_\alpha \sigmav_\gamma \sigmav_\beta
&=& i \epsilon_{\alpha \gamma \beta} {\cal I} \ ,
\end{eqnarray}
where ${\cal I}$ is the unit matrix and $\epsilon_{\alpha \beta \gamma}$
is the antisymmetric tensor. Thus
\begin{eqnarray}
{\bf M}_{\rho \tau} &=&  i \epsilon_{\alpha \gamma \beta}
\left[ \sigmav_\alpha {\bf U}^\dagger {\bf U}
\sigmav_\beta \right]_{\rho \tau} =  i \epsilon_{\alpha \gamma \beta}
[ \sigmav_\alpha \sigmav_\beta]_{\rho \tau}  \nonumber \\
&=& \epsilon_{\alpha \gamma \beta}^2 [\sigmav_\gamma]_{\rho \tau} \ .
\end{eqnarray}
Since this matrix has the same form as before the transformation, we have
shown that the Hubbard Hamiltonian with spin-orbit interactions is invariant
against the transformation by the unitary matrix of Eq. (\ref{TRANS}),
where $\bf U$ is an arbitrary unitary matrix.  Thus the invariance
with respect to arbitrary rotation of pseudospin gives rise to
an arbitrary transformation of real spin.  This indicates that
there is a continuously degenerate manifold for the ground state,
and therefore we do not expect a gap in the excitation spectrum.
A rigorous analysis of the structure of the Goldstone modes
is more complicated\cite{TCL}
and may require a statement of whether or not there
is long-range order in the orbital sector, in order to say whether
these modes are propagating or diffusive.  Intuitively it seems likely that
if there is long-range spin order, there should be propagating spin-wave
modes, at least one of which by our argument will not have an energy gap at
zero wavevector.

One final point is worth noting.  We have discussed that we
expect long-range antiferromagnetic spin ordering in the presence of
spin-orbit coupling.  Once we have antiferromagnetic spin order, the spin-orbit
perturbation acts like a staggered field on the orbital variable, thus
inducing long-range antiferromagnetic order in the orbital variable
$\langle {\bf L} \rangle$.  This type of orbital ordering is not
the same as that in which the thermal expectation value of the
orbital occupation numbers, $\langle N_\alpha \rangle$, become
unequal to one another.

\section{CONCLUSIONS}
In this paper we have analyzed several unusual symmetries of the
three-band $t_{2g}$ Hubbard model and the associated low-energy
KK Hamiltonian for three-fold degenerate orbitals in the ideal ABO$_{3}$
cubic structure. 
As pointed out previously\cite{IHM}, it is evident that
the total number of electrons in any of the three orbital states
is a good quantum number.  In addition,\cite{PRL} the number of electrons
in any arbitrarily chosen plane perpendicular to the $\alpha$-axis
which are in $\alpha$ orbitals is also a good quantum number.
We showed that the Hamiltonian is invariant with
respect to independently rotating the spins of any single orbital
state $\alpha$ for all spins in an arbitrarily chosen plane
perpendicular to the $\alpha$-axis (which is the inactive
axis for such orbital spins).  These symmetries lead to
a dramatic simplification in numerical diagonalizations, as
we illustrated by discussing exact diagonalizations for a cube
of eight sites.  Furthermore, using these symmetries we develop
a rigorous Mermin-Wagner construction, which shows that the
Hubbard model (and perforce the KK Hamiltonian derived from it)
does not support spontaneous long-range spin order at any nonzero
temperature.  These unusual symmetries are destroyed by almost
any realistic perturbation.  
In the presence of spin-orbit coupling, for instance, we
show that a special invariance with respect to rotation of
{\it pseudo-spin} remains unbroken and due to this continuous
symmetry we expect the excitation spectrum to have at least one
gapless Goldstone mode.  In analogy with the situation in
the cuprates\cite{BS,TY1}, the gap one might have expected in
the presence of spin-orbit interactions will only develop in the
presence of both spin-orbit and Coulomb exchange interactions.\cite{MF}  

Of course, in real systems, such as LaTiO$_3$, long-range
spin order and a gap in the elementary excitation spectrum
are observed.\cite{LTO}  However, any credible expression for $T_c$,
for instance, must involve a perturbation beyond the isotropic
``bare'' KK model, because, as we have shown, this model does
not support long-range spin order.  Elsewhere\cite{MF} we
discuss the type of long-range order which mean-field theory
predicts when various perturbations such as a) spin-orbit interactions,
b) nnn hopping, and c) Hund's rule coupling are added
to the KK Hamiltonian. In addition, an important factor to consider
is the distortion and/or rotation of the oxygen octahedra surrounding
each Ti ion.  Similarly, to be credible any theoretical
expression for the gap in the excitation spectrum must involve
perturbations beyond spin-orbit interactions, because we have
shown that with only spin-orbit interactions added to the ``bare''
KK Hamiltonian, the excitation spectrum is gapless.  Finally,
we note that it would be very interesting to study an experimental
system with as small as possible deviations from the 
``bare'' KK model.  Such a system  would show quite exotic properties
(such as peculiar two-dimensional correlations of spins associated
with different orbital flavors).

\newpage
{\bf ACKNOWLEDGMENT}
We acknowledge helpful correspondence with Prof. N. D. Mermin
concerning Ref. \onlinecite{MW}.
ABH thanks NIST for its hospitality during several visits when
this work was done. We acknowledge partial support from the
US-Israel Binational Science Foundation (BSF). The TAU group is
also supported by the German-Israeli Foundation (GIF).




\newpage
\begin{table}
\caption{Hamiltonian matrix for the ground manifold in units of $\epsilon$.
The states are specified by hops relative to the state $|0 \rangle$
shown in Fig. \ref{cube}c. (So $a$ indicates hopping along the $z$-directed
bond labeled $a$, and similarly for $b$, $c$, and $d$.) 
We introduce the notation $\overline \alpha \equiv abcd \alpha$.}
 
\vspace{0.2 in}
\begin{tabular} { || c | c c c c c c c c c c c c c c c c ||} \hline \hline
& 0 & a & b & c & d & ab & bc & bd & ac & ad & cd & $\overline d$ & $\overline c$ &
$\overline b$ & $\overline a$ & $\overline 0$ \\ \hline
0  & $-4$ & $-1 $ & $-1 $ & $-1 $ & $ -1$ & $ 0$ & $ 0$ & 0& 0& $ 0$ & 0& 0& 0& 0& 0& 0\\
a  & $-1 $ & $-2$ & $0 $ & $0 $ & $ 0$ & $ -1$ & $ 0$ & 0& $-1$& $ -1$ & 0& 0& 0& 0& 0& 0\\
b  & $-1 $ & $0 $ & $-2$ & $0 $ & $ 0$ & $ -1$ & $ -1$ & $-1$& 0& $ 0$ & 0& 0& 0& 0& 0& 0\\
c  & $-1 $ & $0 $ & $ 0$ & $-2$ & $0 $ & $ 0$ & $ -1$ & 0& $-1$& $ 0$ & $-1$& 0& 0& 0& 0& 0\\
d  & $-1 $ & $0 $ & $ 0$ & $ 0$ & $-2$ & $ 0$ & $ 0$ & $-1$& 0& $-1$ & $-1$& 0& 0& 0& 0& 0\\
ab & $0 $ & $-1 $ & $ -1$ & $ 0$ & $ 0$ & $-2$ & $ 0$ & 0& 0& $ 0$ & 0& $-1$& $-1$& 0& 0& 0\\
bc & $0 $ & $0 $ & $ -1$ & $ -1$ & $ 0$ & $0 $ & $-2$ & 0& 0& $ 0$ & 0& $-1$& 0& 0& $-1$& 0\\
bd & $0 $ & $0 $ & $ -1$ & $ 0$ & $ -1$ & $0 $ & $ 0$ & 0& 0& $ 0$ & 0& 0& $-1$& 0& $-1$& 0\\
ac & $0 $ & $-1 $ & $ 0$ & $ -1$ & $ 0$ & $0 $ & $ 0$ & 0& 0& $ 0$ & 0& $-1$& 0& $-1$& 0& 0\\
ad & $0 $ & $-1 $ & $ 0$ & $ 0$ & $ -1$ & $0 $ & $ 0$ & 0& 0& $-2$ & 0& 0& $-1$& $-1$& 0& 0\\
cd & $0 $ & $0 $ & $ 0$ & $ -1$ & $ -1$ & $0 $ & $ 0$ & 0& 0& $0 $ &$-2$& 0& 0& $-1$& $-1$& 0\\
$\overline d$
   & $0 $ & $0 $ & $ 0$ & $ 0$ & $ 0$ & $-1 $ & $ -1$ & 0& $-1$&
$ 0$ & $0 $ & $-2$ & $0 $ & $0 $ & $0 $ & $ -1$ \\
$\overline c$
   & $0 $ & $0 $ & $ 0$ & $ 0$ & $ 0$ & $-1 $ & $ 0$ & $-1$& 0&
$ -1$ & $0 $ & $ 0$ & $-2$ & $0 $ & $0 $ & $ -1$ \\
$\overline b$
   & $0 $ & $0 $ & $ 0$ & $ 0$ & $ 0$ & $0 $ & $ 0$ & 0& $-1$&
$ -1$ & $-1 $ & $ 0$ & $0 $ & $-2$ & $ 0$ & $ -1$ \\
$\overline a$
   & $0 $ & $0 $ & $ 0$ & $ 0$ & $ 0$ & $0 $ & $ -1$ & $-1$& 0&
$ 0$ & $-1 $ & $ 0$ & $0 $ & $0 $ & $-2$ & $ -1$ \\
$\overline 0$
   & $0 $ & $0 $ & $ 0$ & $ 0$ & $ 0$ & $0 $ & $ 0$ & 0& 0&
$0 $ & $0 $ & $-1 $ & $-1 $ & $-1 $ & $ -1$ & $-4$ \\ \hline \hline
\end{tabular}
\end{table}

\newpage

\begin{figure}
\includegraphics[scale=3.5]{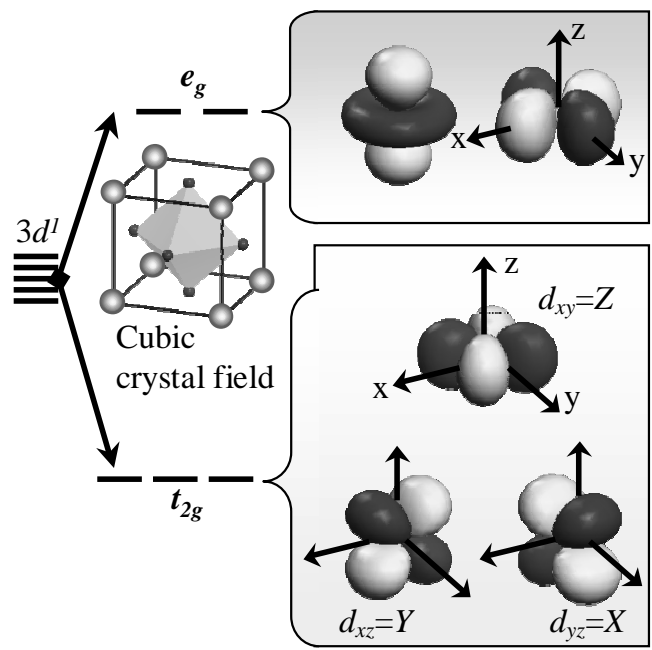}
\bigskip
\bigskip
\bigskip
\caption{A schematic
view of the splitting of the five-fold orbitals under
cubic crystal field. The transition metal is located 
at the center of the oxygen octahedron.}
\label{split}
\end{figure}

\newpage
\begin{figure}

\includegraphics[scale=3.5]{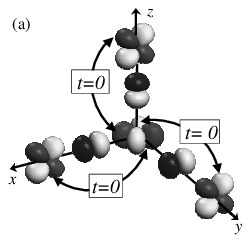}
\bigskip \\
\includegraphics[scale=3.5]{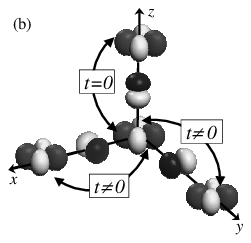}
\bigskip
\bigskip
\bigskip
\caption{Symmetry of the hopping matrix element for cubic site symmetry.
(a) The hopping matrix between different flavors is zero.
(b) The $z$-axis is inactive for $z$ orbitals.}
\label{hop}
\end{figure}

\newpage
\begin{figure}
\includegraphics[scale=3.5]{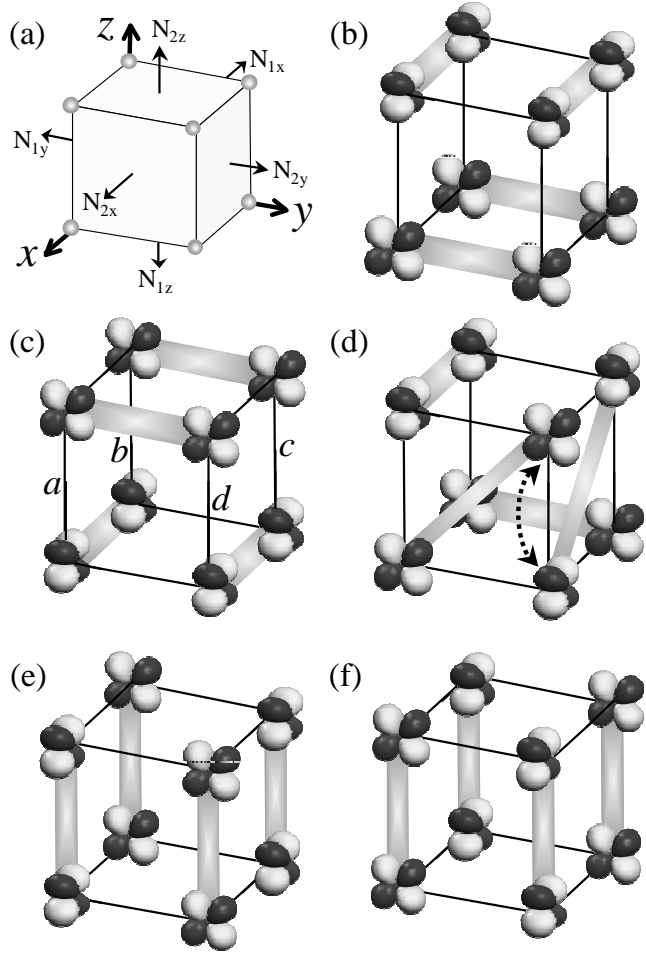}
\bigskip
\bigskip
\caption{Spin and
orbital configurations for a cube of eight sites. The thick lines
indicate singlet spin states (dimers) and $X$ and $Y$ indicate
the orbital states of the electrons. (a) Schematic illustration of
the quantum numbers associated with occupation of $\alpha$ orbitals in
an $\alpha$-plane.  (Similar spin quantum numbers are not shown.)
(b) and (c) Dominant configurations in the ground state wavefunction.
Hopping between $x$ and $y$ orbital states is only allowed along
$z$-direction bonds and these bonds are labeled in panel (c).
(d) A subdominant configuration in the ground state which is obtained
from (c) by allowing the interchange of two ($X$ and $Y$) electrons
along the $z$-axis bond labeled ``d'' while each electron
retains its membership in its original spin singlets (even though the
position of the electron has changed). If $|0\rangle$ denotes
configuration (c), then configuration (b) is
$| \overline 0 \rangle \equiv abcd|0 \rangle$
and (d) is $d|0 \rangle$ in the notation of Table I.
We also show (e) and (f) states with the same quantum numbers 
but which are not coupled to the manifold of 16 states we analyze.
These two states are actually eigenstates of the Hamiltonian
with energy $-4\epsilon$.}
\label{cube}
\end{figure}

\begin{thebibliography} {99}
\bibitem{PERT}
In degenerate perturbation theory we may write the perturbation to the
matrix elements of an operator $A$ due to the perturbation $V$ (which
connects the ground manifold to excited states $|e \rangle$) as 
$\delta \langle n| A| m \rangle
 = - \sum_e [\langle n |A|e \rangle \langle e|V |m \rangle
+ \langle n |V|e \rangle \langle e|A|n \rangle ]/E_e$,
where $|n \rangle$ and $|m\rangle$ are states in the low-energy manifold.
We apply this for $A=c_{{\bf R}+\deltav \alpha \uparrow}^\dagger
c_{{\bf R} \alpha \uparrow}$ and out of the hopping we only need the
term $V=A^\dagger$.  Then, for $E_e=U$ this gives
$\delta \langle n |A| m \rangle = -2 (t/U)\delta_{n,m}$.
\bibitem{CUBE}
There are two additional states (see Figs. \ref{cube}e and \ref{cube}f) 
which also have the quantum numbers
$N_{x,1} =N_{x,2}  =N_{y,1}  = N_{y,2}=2$ and consist of a product of
four spin zero dimers.  In both these states the dimers have their
bonds along the $z$-axis. Such states are not connected by hopping
matrix elements to any states in the manifold of 16 states we consider.
\bibitem{ERROR}
Equation (5) of Ref. \onlinecite{TY1} is misprinted.
The correct version replaces $c^\dagger$ by $c$ and $d^\dagger$ by $d$.
\end{thebibliography}
\end{document}